\documentclass[aps,pra,reprint, longbibliography, twocolumn,superscriptaddress, notitlepage]{revtex4-1}

\usepackage[T1]{fontenc}
\usepackage{lmodern}
\usepackage[utf8]{inputenc}
\usepackage[english]{babel}

\usepackage{indentfirst}

\usepackage{amsthm, amssymb, amstext, amsmath, bbm}
\usepackage{dsfont}

\usepackage{float}
\usepackage{tikz}

\usepackage{epstopdf}
\usepackage{blkarray} 
\usepackage{MnSymbol}

\usepackage{soul}
\usepackage{braket}

\usepackage{xcolor}

\usepackage{hhline}

\usepackage{enumitem}

\definecolor{darkred}{rgb}{0.5,0,0}
\definecolor{darkgreen}{rgb}{0,0.5,0}
\definecolor{darkblue}{rgb}{0,0,0.5}

\usepackage[colorlinks]{hyperref}
\hypersetup{colorlinks, linkcolor=darkred, filecolor=darkgreen,
urlcolor=darkblue, citecolor=darkgreen}

\newcommand{\LL}{\mathcal{L}}
\newcommand{\DD}{\mathcal{D}}

\newcommand{\rhot}{\hat{\rho}(t)}
\newcommand{\de}{{\rm d}}
\newcommand{\sss}{\hat{\rho}_{\rm ss}}
\newcommand{\eig}[1]{\hat{\rho}_{#1}}
\newcommand{\jump}{\hat{\Gamma}}

\renewcommand{\Re}[1]{\mathbb{R}\mathrm{e}\left[#1\right]}

\newcommand{\abs}[1]{\lvert #1 \rvert}

\newcommand{\Tr}[1]{\mathrm{Tr}\!\left[#1\right]}
\usepackage{epstopdf}

\makeatletter
\newcommand*\bigcdot{\mathpalette\bigcdot@{.5}}
\newcommand*\bigcdot@[2]{\mathbin{\vcenter{\hbox{\scalebox{#2}{$\m@th#1\bullet$}}}}}
\makeatother

\newcommand{\setbackgroundcolour}{\pagecolor[rgb]{1,1,1}}
\newcommand{\settextcolour}{\color[rgb]{0.,0.,0}}
\newcommand{\invertbackgroundtext}{\setbackgroundcolour\settextcolour}

\invertbackgroundtext

\newcommand{\invtitle}[1]{\title{\invertbackgroundtext #1}}
\newcommand{\invauthor}[1]{\author{\invertbackgroundtext #1}}
\newcommand{\invaffiliation}[1]{\affiliation{\invertbackgroundtext #1}}
\newcommand{\invdate}[1]{\date{\invertbackgroundtext #1}}

\begin{document}

\invtitle{ Hybrid-Liouvillian formalism connecting exceptional
points of non-Hermitian Hamiltonians and Liouvillians via
postselection of quantum trajectories} \invauthor{Fabrizio
Minganti}\email{fabrizio.minganti@riken.jp }
\invaffiliation{Theoretical Quantum Physics Laboratory, RIKEN
Cluster
    for Pioneering Research, Wako-shi, Saitama 351-0198, Japan}

\invauthor{Adam Miranowicz}\email{adam@riken.jp}
\invaffiliation{Theoretical Quantum Physics Laboratory, RIKEN
Cluster
    for Pioneering Research, Wako-shi, Saitama 351-0198, Japan}
\invaffiliation{Faculty of Physics, Adam Mickiewicz University,
    PL-61-614 Pozna\'n, Poland}

\invauthor{Ravindra W. Chhajlany}\email{ravi@amu.edu.pl}
\invaffiliation{Theoretical Quantum Physics Laboratory, RIKEN
Cluster
    for Pioneering Research, Wako-shi, Saitama 351-0198, Japan}
\invaffiliation{Faculty of Physics, Adam Mickiewicz University,
    PL-61-614 Pozna\'n, Poland}

\invauthor{Ievgen I. Arkhipov} \email{ievgen.arkhipov@upol.cz}
\invaffiliation{Joint Laboratory of Optics of Palack\'y University
and Institute of Physics of CAS, Faculty of Science, Palack\'y
University, 771 46 Olomouc, Czech Republic}

\invauthor{Franco Nori}\email{fnori@riken.jp}
\invaffiliation{Theoretical Quantum Physics Laboratory, RIKEN
Cluster
    for Pioneering Research, Wako-shi, Saitama 351-0198, Japan}
\invaffiliation{Physics Department, The University of Michigan,
Ann
    Arbor, Michigan 48109-1040, USA}

\invdate{\today}

\begin{abstract}
Exceptional points (EPs) are degeneracies of classical and quantum open systems, which are studied in many
areas of physics including optics, optoelectronics, plasmonics, and condensed matter physics. In the semiclassical
regime, open systems can be described by phenomenological effective non-Hermitian Hamiltonians (NHHs)
capturing the effects of gain and loss in terms of imaginary fields. The EPs that characterize the spectra of such
Hamiltonians (HEPs) describe the time evolution of a system without quantum jumps. It is well known that a
full quantum treatment describing more generic dynamics must crucially take into account such quantum jumps.
In a recent paper [F. Minganti \emph{et al}.,
\href{https://journals.aps.org/pra/abstract/10.1103/PhysRevA.100.062131}{Phys.
Rev. A \bf{100}, $062131$ ($2019$)}],we generalized the notion of EPs to the
spectra of Liouvillian superoperators governing open system dynamics described by Lindblad master equations.
Intriguingly, we found that in situations where a classical-to-quantum correspondence exists, the two types of
dynamics can yield different EPs. In a recent experimental work [M. Naghiloo \emph{et
al.} \href{https://www.nature.com/articles/s41567-019-0652-z}
{Nat. Phys. \bf{15}, $1232$ ($2019$)}], it was shown that one can engineer a non-Hermitian Hamiltonian in the quantum limit by postselecting on certain
quantum jump trajectories. This raises an interesting question concerning the relation between Hamiltonian and
Lindbladian EPs, and quantum trajectories. We discuss these connections by introducing a hybrid-Liouvillian
superoperator, capable of describing the passage from an NHH (when one postselects only those trajectories
without quantum jumps) to a true Liouvillian including quantum jumps (without postselection). Beyond its
fundamental interest, our approach allows to intuitively relate the effects of postselection and finite-efficiency
detectors.
\end{abstract}

\maketitle

%
%


\tableofcontents

\section{Introduction}

Exceptional points (EPs) are at the center of intense theoretical and experimental research in many different areas of
physics, such as optics, condensed matter physics, plasmonics, and even electronics. Originally, EPs were studied in connection to parity-time ($\mathcal{PT}$) symmetric non-conservative
systems~\cite{Bender1998,ChristodoulidesBOOK,PengScience14}. Since an EP
corresponds to a non-diagonalizable operator, standard Hermitian
Hamiltonians cannot display any EP. It is the non-unitary action
of the environment on the system that induces the emergence of
EPs. Such points have been studied by balancing attenuation,
amplification, gain saturation, as well as various Hamiltonian
coupling strengths of an open system (for experimental
realizations, see e.g.,
Refs.~\cite{PengScience14,Peng2014,GaoNature15,NaghilooNatPhys19}).

The interest in studying EPs, however, is not limited to
$\mathcal{PT}$-symmetric systems. Indeed, in proximity to other
types of EPs, a system can display exotic phenomena, such as EP-induced lasing~\cite{JingPRL14,PengScience14,ZhangNaturePhotonics2018} or even new forms of photon blockade~\cite{HuangarXiv20}. Beyond the purely phenomenological interest in studying these systems, EPs are considered for novel apparatus
harnessing peculiar properties of mode coalescence at an EP~\cite{Ozdemir2019,Miri2019}.

Many results obtained so far focused on non-Hermitian
Hamiltonians (NHHs), i.e., systems in which losses, gain,
and its saturation, decoherence, etc. are considered only as
imaginary-valued fields. An EP of an NHH (which for brevity
we refer to as a Hamiltonian EP or an HEP) refers to those
NHH degeneracies where two (or more) eigenfrequencies coincide and the corresponding eigenstates coalesce. Arguably,
one can categorize those systems as (semi)classical, since the
effect of the environment has not been taken into account
according to a full-quantum description.

Recently, the quest for true quantum EPs has attracted
much attention \cite{HatanoMP19,NaghilooNatPhys19,MingantiPRA19,PerinaPRA19,ArkhipovPRA20,HuangarXiv20,KuoPRA20,JaramilloScirep20}. 
An open \emph{quantum} system interacting with its environment must include dissipative terms
describing the progressive loss of energy, coherence, and
information transfer to the environment  \cite{Gardiner_BOOK_Quantum,Haroche_BOOK_Quantum,Carmichael_BOOK_1,Wiseman_BOOK_Quantum,BreuerBookOpen}. For a weakly
coupled Markovian environment, the (Gorini-Kossakowski-Sudarshan-Lindblad, or, for the sake of brevity) Lindblad
master equation can efficiently capture the dynamics of the system. 
The Lindblad master equation consists of a Hermitian
Hamiltonian part (i.e., the coherent evolution of the system)
and the so-called Lindblad dissipators, which characterize the
coupling with the environment \cite{LidarLectureNotes}. These Lindblad dissipators can, in turn, be divided into two parts: the first one represents a
\emph{coherent nonunitary dissipation} of the system, similar
to the imaginary-frequency fields of an NHH; the second
part describes \emph{quantum jumps}, which result from the effect
of a continuous measurement performed by the environment
on the system~\cite{Wiseman_BOOK_Quantum, Haroche_BOOK_Quantum,
Barnett_BOOK_Info, ParisEPJST2012}. Quantum jumps induce both a
random and instantaneous change of the stochastic wave function describing the system and a continuous non-Hermitian
dynamics of the system~\cite{MolmerJOSAB93, Haroche_BOOK_Quantum,
Carmichael_BOOK_2, Wiseman_BOOK_Quantum,
DaleyAdvancesinPhysics2014}.  Such instantaneous
switching caused by quantum jumps is fundamental to obtain
a theory of the system-environment interaction consistent
with a probabilistic interpretation of quantum-mechanical
measurements~\cite{Wiseman_BOOK_Quantum,
DaleyAdvancesinPhysics2014, Carmichael_BOOK_1, Carmichael_BOOK_2}.
Moreover, such quantum jumps
have been experimentally observed in many setups~\cite{NagourneyPRL86, SauterPRL96,BergquistPRL99,
PeilPRL99,BascheNat95,GleyzesNat07, GuerlinNat07,DelegliseNat08,
SayrinNat11, JelezkoAPL02,NeumannScience10,
RobledoNat11,VijayPRL11, HatridgeScience13, SunNature2014,
OfekNat16, Minev2019}.

To every Lindblad master equation corresponds a nonHermitian Liouvillian superoperator $\LL$ which can display Liouvillian exceptional points (LEPs)~\cite{LidarLectureNotes, AlbertPRA14,
MingantPRA18_Spectral, MacieszczakPRL16, SarandyPRA05_Adiabatic,
HatanoMP19, ProsenJST10,MathisenENTROPY2018, MoosSciPost19}. LEPs are
defined via degeneracies of Liouvillians (i.e., including the
effects of quantum jumps) as introduced in Ref.~\cite{MingantiPRA19} where
their connection with HEPs has also been investigated (see
also Refs.~\cite{ArkhipovPRA20,HuangarXiv20}).

It has been noticed that an NHH naturally emerges when
discussing \emph{quantum trajectories} and \emph{postselection}~\cite{NaghilooNatPhys19}. 
As mentioned above, quantum trajectories describe a system whose environment is continuously and perfectly probed. 
 As
mentioned above, quantum trajectories describe a system
whose environment is continuously and perfectly probed.
Even if quantum jumps cannot be avoided, if one postselects
only those trajectories where no quantum jumps took place,
the effective resulting dynamics is that of an NHH. For
instance, in Ref.  \cite{NaghilooNatPhys19}, such a postselection was used to explore
the properties of the NHH of a superconducting three-level
system.

From a Liouvillian point of view, postselecting the trajectories without quantum jumps corresponds to consider a
Liouvillian without quantum jumps ($\LL'$) \cite{MingantiPRA19}.
Even if a perfect
postselection of all quantum jumps is captured by $\LL'$, this ideal
case may not be always realizable. For instance, one may be
not able to postselect all the quantum jumps characterizing the
system. Moreover, no real instrument can perfectly monitor
the system, and therefore a perfect postselection is impossible.
This motivates the question of how EPs and their associated
effects depend on these protocols.

The main objective of this paper is to answer this question,
addressing the relationship between HEPs, LEPs, and imperfectly postselected trajectories. For this, we introduce a hybrid
Liouvillian
$\LL_{H}(q)$, a generalization of $\LL$ and $\LL'$, capable of
describing those (imperfect) processes. Roughly speaking, the
quantum-jump parameter  $q \in [0,1]$ describes how much one allows quantum jumps to
affect the dynamics of a density matrix. This very formal definition allows to relate and compare
HEPs with LEPs, and the corresponding evolutions between
these two limits, i.e., the classical-to-quantum transition of
EPs. By considering the protocol of quantum trajectories, we
demonstrate that the hybrid Liouvillian $\LL_{H}(q)$ has a very clear
and specific physical meaning as the average over only a
certain type of postselected trajectories. Indeed, $\LL_{H}(q=0)$ represents a perfectly monitored system where we postselect only those trajectories where no quantum jump occurred (thus
recovering an NHH). On the contrary, $\LL_{H}(q=1)$ describes
the average of trajectories where no postselection has been
applied, thus recovering a full Liouvillian description. Finally,
we attribute the case $0<q<1$ to a postselected system in the
presence of an imperfect measurement instrument.

This paper is organized as follows. In Sec. \ref{Sec:Liouv_and_quant_traj}, we recall
the basic elements of an open-system description in terms of
Lindblad operators, define LEPs and HEPs, recall the unraveling of the master equation in terms of quantum trajectories,
and summarize postselection. In Sec. \ref{Sec:Hybrid-Liouvillian_post-selection}, we introduce the
hybrid-Liouvillian formalism and provide two interpretations
for it. In Secs. \ref{Sec:Example_1}~and~\ref{Sec:Example_2} we discuss instructive examples of
the classical-to-quantum transition of EPs. Finally, we discuss
the implications of our results in the Conclusions. In the Appendix~\ref{App:eigvals}
we also provide the analytical expressions for the
EPs in one of the studied examples of the hybrid Liouvillian.

In the main article, we will use all the previously introduced abbreviations. 
In Table \ref{Tab:Abb} we concisely list them, to facilitate the reading of the article. 

\begin{table}
	\begin{tabular}{c|c}
		Full Name & Abbreviation \\
		\hhline{=|=}
		Non-Hermitian Hamiltonian     &  NHH\\
		\hline
		Liouvillian     & $\LL$ \\
		\hline
		Liouvillian without quantum jumps     & $\LL'$ \\
		\hline
		Exceptional point & EP \\
		\hline
		Hamiltonian exceptional point & HEP
		\\
		\hline
		Liouvillian exceptional point & LEP \\
		\hline Hybrid Liouvillian & $\LL_{H}(q)$ \\
		\hline Quantum jump parameter & $q$
	\end{tabular}
	\caption{List of abbreviation used in the main article, and corresponding full names.}
	\label{Tab:Abb}
\end{table}

\section{Liouvillians and Quantum Trajectories}
\label{Sec:Liouv_and_quant_traj}

An open quantum system weakly interacting with a Markovian (i.e., memoryless) environment can be described using a
Lindblad master equation:
\begin{equation}\label{Eq.Linblad}
\partial_t \rhot = - i [\hat{H}, \rhot] + \sum_{\mu} \DD[\jump_\mu] \rhot.
\end{equation}
Here, $\rhot$ represents the density matrix of the system, that
is, the operator encoding all the information available to an
external observer without any knowledge (or control) of how
the environment \emph{microscopically and instantaneously} acts on the
system. The operator $\hat{H}$ is the Hamiltonian, describing the
coherent evolution of the system, while $\jump_\mu$ are quantum jump
operators, describing the overall effect of the environment on
the dynamics of the system. The jump operators act via the
Lindblad dissipators as
\begin{equation}
\DD[\jump_\mu] \rhot= \jump_\mu \rhot \jump_\mu^\dagger -
\frac{1}{2}\left[\jump_\mu^\dagger \jump_\mu \rhot + \rhot
\jump_\mu^\dagger \jump_\mu\right]\,.
\end{equation}

The effect of $\DD [\hat{\Gamma}_\mu]$ on the density matrix
$\rhot$ can be split into two parts~\cite{Wiseman_BOOK_Quantum}:
the \emph{continuous} non-unitary dissipation terms,
$\hat{\Gamma}_\mu^\dagger \hat{\Gamma}_\mu \rhot + \rhot
\hat{\Gamma}_\mu^\dagger \hat{\Gamma}_\mu$, and the \emph{quantum
jump} terms $\hat{\Gamma}_\mu \rhot \hat{\Gamma}_\mu^\dagger$.
This dissipation describes the continuous
and small loss of energy, information, and coherence of the system into the environment. The quantum jumps describe the
abrupt changes of the state of the system due to dissipation,
and can be thought of as the measurementlike action of a
macroscopic environment on the state of the system
\cite{Wiseman_BOOK_Quantum,Haroche_BOOK_Quantum}.

Equation~\eqref{Eq.Linblad} is linear in $\rhot$ and,
consequently, one can associate a superoperator to it --  the so-called
Liouvillian superoperator $\LL$ \cite{LidarLectureNotes}. This is a superoperator in the sense that it
acts on an operator (the density matrix) to produce an operator
analogous to the way in which an operator acts on a vector
to produce a new vector~\cite{Carmichael_BOOK_2}. Using the Liouvillian $\LL$, we
have
\begin{equation}\label{Eq:Liouvillian}
\partial_t \rhot= \LL \rhot.
\end{equation}
Hereafter, we assume that $\LL$ is time
independent.

\subsection{Liouvillian spectrum, Hamiltonian and Liouvillian EPs}

The spectrum of the Liouvillian $\cal{L}$ is found according to
the formula
\begin{equation}\label{Eq:spectrum}
{\cal{L}}\hat\rho_i=\lambda_i\hat\rho_i,
\end{equation}
where $\lambda_i$ and $\hat\rho_i$ are the eigenvalues and
eigenmatrices of the Liouvillian, respectively. For convenience,
we sort the eigenvalues in such a way that
$\abs{\Re{\lambda_0}}<\abs{\Re{\lambda_1}} < \ldots <
\abs{\Re{\lambda_n}}$. 

The models which we consider are finite dimensional
and time independent. In this regard, one is guaranteed that
there always exists a zero eigenvalue of the Liouvillian. This
represents the fact that there exists at least one steady state $\sss$ towards which the system evolves. Even if the steadystate density matrix associated with the zero is important in
many aspects of open quantum systems, in this case it is not
particularly interesting since it cannot display any exceptional
point (see the discussion in Ref. \cite{MingantiPRA19}).

For a Liouvillian with a unique steady state, the eigenmatrix $\eig{0}$, associated with $\lambda_0=0$,
defines the steady-state density matrix $\hat\rho_{\rm
ss}\propto\hat\rho_0$ of the system. The eigenmatrices $\eig{i}$,
with $i>0$,  describe the transient dynamics towards the steady
state. For a more detailed discussion of the properties of the
Liouvillian spectrum, see Refs.~\cite{LidarLectureNotes,AlbertPRA14,MacieszczakPRL16,MingantPRA18_Spectral,MingantiPRA19}.

In this formalism, LEPs describe the coalescence of two
eigenmatrices of the Liouvillian, for some appropriate choice
of parameters. At an EP, the Liouvillian is defective and cannot be diagonalized. For a LEP of order 2, the eigenvalue $\lambda_j$ admits only one  eigenmatrix $\eig{j}$.
However, one can introduce a generalized eigenmatrix $\hat{\tilde{\rho}}_j$, which is defined via the Jordan chain:
\begin{equation}\label{Eq:Jordan_chain}
\LL \hat{\tilde{\rho}}_j = \lambda_j \hat{\tilde{\rho}}_j + \eig{j}.
\end{equation}
This generalized eigenmatrix completes the basis of the other eigenmatrices $\eig{i}$, i.e., any operator
can be written as a linear combination of the $\eig{i}$ and $\hat{\tilde{\rho}}_j$.
In this basis, the Liouvillian attains its Jordan canonical form.

One of the central results proved  in Ref.~\cite{MingantiPRA19} is
that LEPs should be understood as \emph{purely dynamical} phenomena since
in this Lindblad ME formalism, LEPs can emerge only for the
``excitations''  $\eig{i}$ ($i>0$) above the steady state $\eig{0}$.

Given a Lindblad master equation as in Eq.~\eqref{Eq.Linblad}, we can
introduce the corresponding effective NHH of the form
\begin{equation}\label{Eq:Eff_Ham_1}
\hat{H}_{\rm eff} = \hat{H} - i \sum_{\mu}
\frac{\hat{\Gamma}_\mu^\dagger \hat{\Gamma}_\mu}{2}.
\end{equation}
Note that $\hat{H}_{\rm eff}$ is a NHH since $\hat{H}_{\rm
eff}^\dagger \neq \hat{H}_{\rm eff}$. The equation of motion for a
generic density matrix $\rhot$, thus, becomes
\begin{equation}\label{Eq:Linblad_semiclassical}
\frac{\partial \hat{\rho}(t)}{\partial t} = \LL  \hat{\rho}(t)= -
i \left[\hat{H}_{\rm eff}\hat{\rho}(t) -  \hat{\rho}(t)
\hat{H}_{\rm eff}^\dagger \right] + \sum_{\mu} \jump_{\mu} \rhot
\jump_{\mu}^\dagger.
\end{equation}
If one assumes  that the effect of the jump operators $\sum_{\mu}
\jump_{\mu} \rhot \jump_{\mu}^\dagger$ is always zero during this
evolution, the evolution of the system is provided by
$\hat{H}_{\rm eff}$. Such non-Hermitian operators may support
EPs in their spectra, which we refer to here as NHH Hamiltonian
EPs (HEPs) in contrast to LEPs.

\subsection{Quantum trajectories}

From a theoretical point of view, there are two very different physical interpretations which can be associated with
the Lindblad master equation. The first interpretation is to
consider that a true action of the environment on the system is
impossible to be known exactly, so that the dissipators $\DD[\jump_{\mu}]$ describe the average effect of the environment. In this sense,
the density matrix $\rhot$ is a statistical mixture since one does
not know the details of the system-environment interaction.

On the contrary, if we were to know perfectly the action
of the system on the environment, then we could model it
as a series of perfect measurement instruments \cite{Haroche_BOOK_Quantum,Gardiner_BOOK_Quantum,Wiseman_BOOK_Quantum}. 
In this description, the action of the dissipators $\DD[\jump_\mu]$is to
induce random changes in the system (associated with the
detection of one of the operators
$\jump_{\mu}$). Once an average over
several realizations of the same protocol is considered, the
randomness associated with the dissipator action introduces
a statistical mixture of pure states, resulting in the density
matrix $\rhot$.

Both approaches lead to the same average results
\cite{Haroche_BOOK_Quantum,Wiseman_BOOK_Quantum}; i.e., the average descriptions of the system evolution are
equivalent. However, while according to the first interpretation it is conceptually difficult to consider the state of a
quantum system during a single experimental realization, the
second approach allows to describe an idealized evolution of
the system whose environment is continuously and perfectly
probed (or monitored). Such an equation of motion is called a \emph{quantum trajectory} (for a more detailed discussion, see,
e.g., Refs.~\cite{MolmerJOSAB93,CarmichaelPRL93,
Carmichael_BOOK_2, DaleyAdvancesinPhysics2014,DalibardPRL92}). 
 In this formalism, the state of the
system along a trajectory is described by a wave function $\ket{\psi(t)}$,  which evolves stochastically, and the results of the
Lindblad master equation are recovered by averaging over
many trajectories.

Theoretically, the simplest measurement instrument continuously monitoring the environment is the one which produces only two outcomes: one if the desired state is detected,
zero otherwise. We can imagine that, for each dissipator, there
is an instrument measuring if a quantum jump takes place
continuously, perfectly, and instantaneously. By counting the
number of quantum jumps which are taking place of each
type, we can reconstruct the state of a given system~\cite{Wiseman_BOOK_Quantum}. Using this
\emph{trajectory counting} apparatus for an infinitesimal time
$\de t$, the evolution of the wave function can be described by:
\begin{equation}
\label{Eq:SSE}
\begin{split}
\de \ket{\psi(t)}= & \left[ \sum_{\mu} \de N_{\mu}(t) \left(\frac{\jump_{\mu}}{\sqrt{\braket{\jump^\dagger_{\mu} \jump_{\mu} }}} - \mathds{1}\right) \right. \\
 & \quad \left.  + \de t \left( - i   \hat{H}_{\rm eff} + \sum_{\mu}    \frac{\braket{\jump^\dagger_{\mu} \jump_{\mu} }}{2} \right)\right]\ket{\psi(t)},
\end{split}
\end{equation}
where the effective Hamiltonian is that introduced in Eq.~\eqref{Eq:Eff_Ham_1}.

Note that $\de \ket{\psi(t)}$  is a differential describing either the abrupt evolution
with a stochastic quantum jump [the term proportional to $\de N_\mu (t)$] or the smooth nonunitary evolution dictated by $\hat{H}_{\rm eff}$. 
The expectation values of the jump operators, instead, ensure that
the wavefunction $\de \ket{\psi(t)}$ remains well-normalized along the dynamics.
For a more detailed discussion, see Eq.~(4.17) of Ref. \cite{Wiseman_BOOK_Quantum}.

The  stochastic ``counting'' parameters $ N_\mu(t)$ contain the
information about the total number of jumps which took place along
the dynamics from the initial time $t=0$ to time $t$. Hence, $\de
N_{\mu}(t)= N_\mu(t+\de t) - N_\mu(t)$ is the dichotomic stochastic
variable representing the detection outcome at time $t$. 
Specifically, $\de N_{\mu}(t)=0$ [$\de N_{\mu}(t)=1$] if no (one)
quantum jump $\jump_{\mu}$ took place. 
Hence, one cannot simply take the derivative of $\de N(t)$ with respect to $\de t$.
One can, however, define the probability that a
quantum jump occurs during a time $\de t$ as
\begin{equation}
p\left[\de N_{\mu}(t)=1\right]=\braket{\jump_{\mu}^\dagger
\jump_{\mu}} \de t.
\end{equation}

The NHH can thus be interpreted as the operator determining the dynamics between two successive quantum jumps.
Furthermore, the terms $\braket{\jump^\dagger_{\mu}
\jump_{\mu} }$ in Eq.~\eqref{Eq:SSE} act as normalization
constants necessary to ensure that $\braket{\Psi(t)|\Psi(t)}=1$.
Equivalently, we can think that the system evolves under the action
of $\hat{H}_{\rm eff}$ and we have to renormalize the wave function at each
time step. The above interpretation allows also for simple
efficient Monte Carlo simulation of the ensuing dynamics~\cite{MolmerJOSAB93, Haroche_BOOK_Quantum,
Wiseman_BOOK_Quantum, qutip1, qutip2}.


Finally, we note that the trajectory-counting-based monitoring of the operators
$\jump_{\mu}$ is not the only possible unraveling
of the master equation. Indeed, there exist other possible
choices of jump operators which result in the same master
equation once the average over many quantum trajectories
is taken~\cite{Haroche_BOOK_Quantum, Wiseman_BOOK_Quantum}.
 Different unraveling can result in extremely
different dynamics at a single trajectory level \cite{BartoloEPJST17,RotaNJP18,Haroche_BOOK_Quantum}.
In
this sense, the use of a Lindblad master equation allows to
capture those properties which do not depend on the details of
the system-environment exchange.

\subsection{Postselection of quantum trajectories}

Suppose now that we observe a quantum trajectory where, at time
$t$, $ N_\mu(t)=0$ for all $\mu$. We conclude that the system has
evolved under the genuine action of the NHH $\hat{H}_{\rm eff}$ in
Eq.~\eqref{Eq:SSE}.  In this regard, by postselecting the trajectory with no
quantum jumps (i.e., discarding all those which do present
some quantum jumps) one can obtain an NHH also in the
quantum limit, that is, when normally quantum jumps would play a fundamental role in correctly describing the physics
of the system. As has been shown in Ref.~\cite{NaghilooNatPhys19},  this procedure
allows to study the emergence of HEPs also in quantum
systems.

There are, however, some necessary remarks concerning
this postselection procedure. First, we notice that, in this way,
we cannot experimentally connect many-particle semiclassical EPs to the fully quantum ones. Indeed, in the semiclassical
limit of a many-particle system, many quantum jumps must
happen, and the probability to observe a trajectory without
quantum jumps rapidly tends to zero.

Indeed, postselecting in the semiclassical limit would be
equivalent to avoiding environment-induced superselection
(einselections), collapsing the ``quantum'' wave function into
a classical state  \cite{ZurekRMP03}. 
For example, in an optical cavity with jump operator $\hat{a}$ ($\hat{a}$ being the bosonic destruction operator), the number of jumps per unit of time $\de t$ is roughly given by $\de  t \cdot \braket{\Psi(t)| \hat{a}^\dagger \hat{a}| \Psi(t)}$.
In a many-particle system, this number is extremely high, making it almost impossible to observe a trajectory without quantum jumps.

Moreover, to truly observe a HEP it is necessary to have a
perfect measurement instrument which collects all the quantum jumps and that never allows a quantum jump to go undetected. Hence, in principle, true postselection is impossible,
which leads to two questions:

(i) How can we relate the results of NHHs and Liouvillians
in a more formal way?

(ii) How can we describe the consequences of imperfect
monitoring, i.e., finite efficiency detectors in quantum trajectories?

\section{Hybrid-Liouvillian formalism and its connection to postselection}
\label{Sec:Hybrid-Liouvillian_post-selection}
To answer both questions raised in the previous section, we
introduce a hybrid-Liouvillian formalism; i.e., we introduce a
modified Liouvillian superoperator. To better understand this
hybrid Liouvillian, here we focus on Eq.~\eqref{Eq:Liouvillian} which in the case
where there is just one quantum jump becomes
\begin{equation}\label{Eq:Liouvillian_superoperator_dot_form}
\LL = - i \left[\hat{H}, \bigcdot\right] + \jump \bigcdot
\jump^\dagger - \frac{\jump^\dagger\jump \bigcdot + \bigcdot
\jump^\dagger\jump}{2},
\end{equation}
where $\bigcdot$ is a placeholder to indicate where an operator should be applied.
As already mentioned, ignoring  the effect of quantum jumps $\jump
\bigcdot \jump^\dagger$, one obtains a non-Hermitian Hamiltonian
evolution which can be recast in superoperator form as
\begin{equation}
\LL'=  - i  \left( \hat{H} - i \frac{\jump^\dagger\jump}{2}
\right) \bigcdot + i \bigcdot \left( \hat{H} - i
\frac{\jump^\dagger\jump}{2} \right)^\dagger = -i \hat{H}_{\rm
eff} \bigcdot + i \bigcdot \hat{H}_{\rm eff}^\dagger.
\end{equation}
The superoperator $\LL'$ is the Liouvillian without quantum jumps, and its spectrum is fully determined by that of $\hat{H}_{\rm eff}$ \cite{MingantiPRA19}.
This equation is not trace preserving.  This problem is
solved as in the case of evolution with $\hat{H}_{\rm eff}$ in a
quantum trajectory, where  the density matrix can be renormalized
at each time step to ensure that $\Tr{\rhot}=1$. Now, by not
completely ignoring the effects of quantum jumps, one can
formally introduce a hybrid Liouvillian of the form:
\begin{equation}
\begin{split}
\LL_{\rm H}(q) &=- i \left[\hat{H}, \bigcdot\right] + q \, \jump \bigcdot \jump^\dagger - \frac{\jump^\dagger\jump \bigcdot + \bigcdot \jump^\dagger\jump}{2} \\
&= - i \left[\hat{H}, \bigcdot\right] + \DD[\sqrt{q}\jump] -
\left(1-q\right) \frac{\jump^\dagger\jump \bigcdot + \bigcdot
\jump^\dagger\jump}{2}.
\end{split}
\end{equation}

The hybrid Liouvillian $\mathcal{L}_{\rm H}(q)$ has a clear
mathematical significance, viz., it is the weighted average of $\LL$ and $\LL'$. Clearly it interpolates
between NHH evolution ($q=0$) to a completely Liouvillian one
($q=1$) and the transition of EPs between these two limits can
be traced by tuning the quantum jump parameter $q$. The physical meaning of
this hybrid operator is explained below.

\subsection{Interpretation of the hybrid Liouvillian in terms of post-selected trajectories}

\begin{figure}
    \centering
    \includegraphics[width=0.7\linewidth]{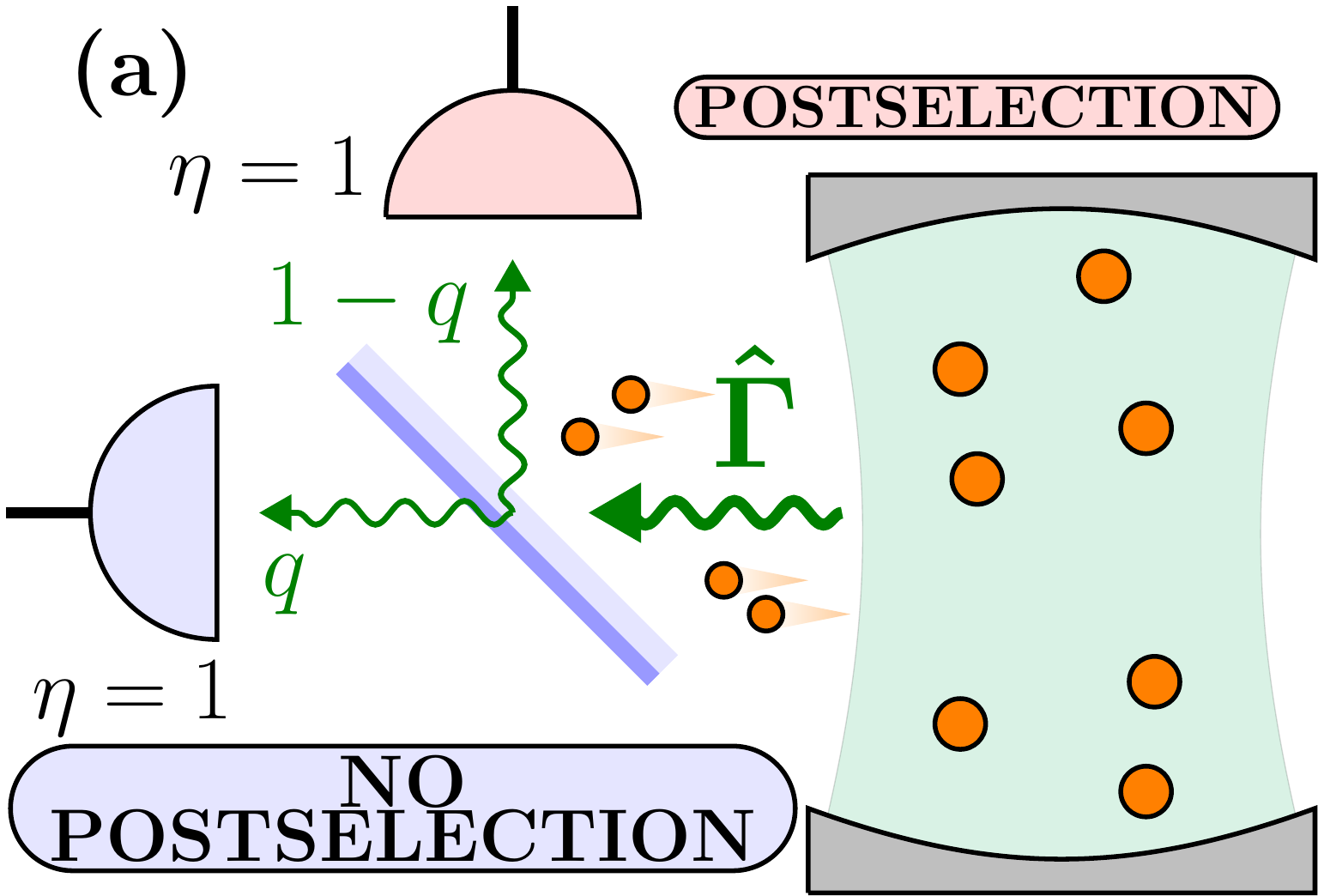}\vspace{0.5cm}
    \includegraphics[width=0.7\linewidth]{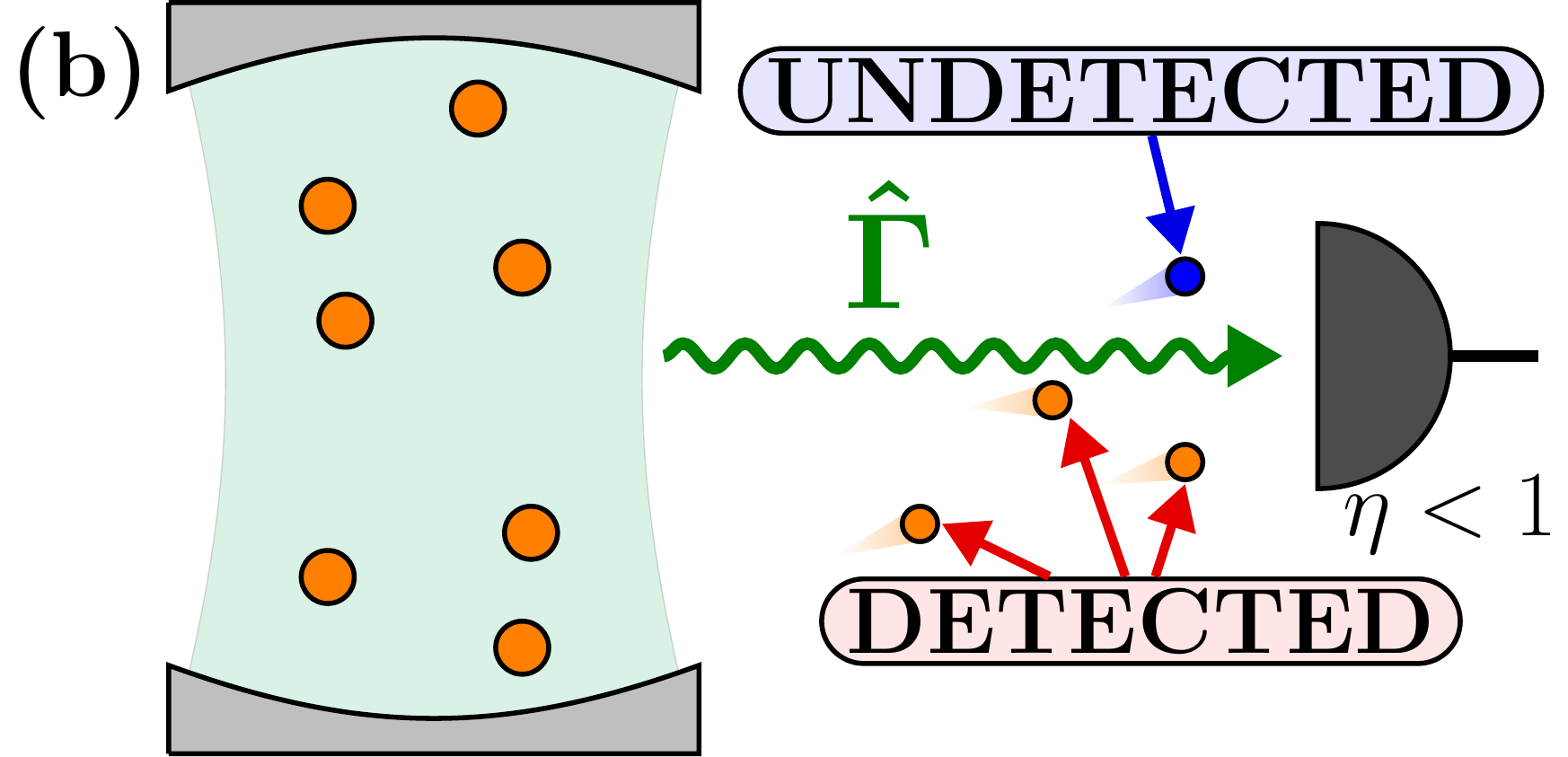}
\caption{Pictorial representation of the physical systems for which
	evolution is described by the hybrid Liouvillian $\LL_H (q)$
depending on
the quantum jump parameter $q$.
To clarify the ideas, we sketch an
optical cavity with one-photon loss as dissipator (the jump operator $\jump$).  (a) A system with two perfect photodetectors (efficiency $\eta=1$)
simultaneously measuring the same jump operator $\jump$
for a photon
leaking from the cavity and passing through a beam splitter (BS) with
the transmission (reflection) probability $q$ ($1-q$). 
Thus, the quantum
jump parameter $q$
corresponds to the probability of measuring a
leaking photon by the left photodetector. According to this setup,
we postselect only those trajectories which read $N^{(2)}(t)=0$; that
is, no quantum jumps were detected in the photodetector (detecting
photons with probability $1-q$, the upper one in the drawing). 
(b) A finite-efficiency photodetector ($\eta<1$).While a certain fraction of
the photons will be detected (the orange ones), some of them will
not (the blue one). We then perform a postselection requiring that
no quantum jumps were detected. Since not all the photons which
escaped the system where detected, we may average also over some
quantum trajectories where a photon escaped the system. This hybrid
Liouvillian describing this system is $\LL_H(q=1-\eta)$.
    }
    \label{fig:twodetectors}
\end{figure}

The Liouvillian in
Eq.~\eqref{Eq:Liouvillian_superoperator_dot_form} can be
conveniently recast as:
\begin{equation}\label{Eq:Liouvillian_broken_0}
\begin{split}
\LL &= - i \left[\hat{H}, \bigcdot\right] + \DD[\jump]=- i \left[\hat{H}, \bigcdot\right] + q \DD[\jump]+ (1-q) \DD[\jump] \\
&= - i q \left[\hat{H}, \bigcdot\right] +  \DD[\sqrt{q} \, \jump]
- i (1-q) \left[\hat{H}, \bigcdot\right] + \DD[\sqrt{1-q}
\,\jump].
\end{split}
\end{equation}
From a quantum trajectory point of view,
Eq.~\eqref{Eq:Liouvillian_broken_0} means that
instead of having only one measuring instrument collecting
the jumps of the operator
$\jump$, we have two different detectors [c.f.
Fig.~\ref{fig:twodetectors} (a)], so that
\begin{equation}
\begin{split}\label{Eq:counting_with_two_noises}
\de \ket{\psi(t)} &=\left\{\rule{0cm}{0.8cm} \left[ \left(\rule{0cm}{0.35cm}\de N^{(1)}(t) + \de N^{(2)}(t)\right)\left(\frac{\jump}{\sqrt{\braket{\jump^\dagger \jump }}}  - \mathds{1}\right)\right] \right.\\
& \qquad - i  \de t \left(\hat{H}_{\rm eff}+ \frac{\braket{\jump^\dagger
\jump}}{2}\right) \left. \rule{0cm}{0.8cm} \right\} \ket{\psi(t)},
\end{split}
\end{equation}
where $\hat{H}_{\rm eff}$ is
\begin{equation}\label{Eq:H_eff_two_detectors}
\begin{split}
\hat{H}_{\rm eff} & =\hat{H}_{\rm eff}^{(1)}+\hat{H}_{\rm eff}^{(2)}, \\
\hat{H}_{\rm eff}^{(1)}&=\hat{H} - i q  \frac{\jump^\dagger \jump}{2},\\
\hat{H}_{\rm eff}^{(2)}&=\hat{H} - i (1-q)  \frac{\jump^\dagger \jump}{2}. \\
\end{split}
\end{equation}
The total probability of a jump in an infinitesimal amount of time
is given by
\begin{equation}\label{Eq:proba_tot_two_detectors}
p\left[\de N^{(1)}(t)+\de
N^{(2)}(t)=1\right]=\braket{\jump^\dagger \jump} \de t.
\end{equation}
and
\begin{equation}\label{Eq:proba_two_detectors}
\begin{split}
p\left[\de N^{(1)}(t)=1\right] &= q \braket{\jump^\dagger \jump} \de t, \\
 p\left[\de N^{(2)}(t)=1\right] &= (1-q) \braket{\jump^\dagger \jump} \de t.
\end{split}
\end{equation}
That is,
Eqs.~\eqref{Eq:counting_with_two_noises},~\eqref{Eq:H_eff_two_detectors},~and~\eqref{Eq:proba_two_detectors}
correspond to assuming that a fraction $q$ of detections will happen in the first detector, while
all the other detections happen in the second one. The average
total number of detections in an experiment, however, must
be identical to a setup where only one detector is present [cf.
Eq.~\eqref{Eq:proba_tot_two_detectors}].

Let us assume that we postselect the results of the second detector; i.e., we choose only those trajectories where
no quantum jump took place for the second detector and
$N^{(2)}(t)=0$. Detector 1 will exactly produce the Lindblad
master equation with dissipator $\DD[\sqrt{q}\jump]$,  while the second
detector will produce a time evolution via its non-Hermitian
Hamiltonian $\hat{H}_{\rm eff}^{(2)}$. Hence
the Liouvillian in Eq.~\eqref{Eq:Liouvillian_broken_0} becomes
\begin{equation}
- i \left[\hat{H}, \bigcdot\right] +  \DD[\sqrt{q} \, \jump]  -
\left(1-q\right) \frac{\jump^\dagger\jump \bigcdot + \bigcdot
\jump^\dagger\jump}{2} \equiv \LL_{H}(q).
\end{equation}

We have therefore proved that $\LL_{H} (q)$ describes the
evolution of the state monitored by two perfect instruments, one
of which is post-selected. Hence, $\LL_{H} (q)$ is a physically
legitimate quantum map. Note that one must, however, renormalize the
density matrix to ensure that its trace is 1.

\subsection{ Postselected quantum trajectory and inefficient detectors}

We can also assign a different, experimentally relevant meaning to
Eq.~\eqref{Eq:counting_with_two_noises}. Let us consider a
finite-efficiency detector, such that, with probability $\eta$,
a quantum jump happens but the detector does not report it happening. The
Lindblad master equation of this system is (see, e.g., page 190 of
Ref.~\cite{Wiseman_BOOK_Quantum}):
\begin{equation}\label{Eq:Liouvillian_broken}
\begin{split}
\LL(\eta) &= - i \left[\hat{H}, \bigcdot\right] + \DD[\jump]=- i
\left[\hat{H}, \bigcdot\right] + \eta \DD[\jump]+ (1-\eta)
\DD[\jump].
\end{split}
\end{equation}
Again, we can model such a system as one in which we
have two perfect detectors which continuously monitor the
system and collect all the quantum jumps which take place.
Even if, theoretically, the effects of these two detectors on a
single quantum trajectory are identical to the presence of a
single detector, the description is extremely different once we
try to postselect the results. Indeed, one of the two detectors
does not pass any information to an observer, which cannot
know if a quantum jump took place. The description of such
a system is, therefore, exactly captured by the hybrid  Liouvillian
$\LL_{H}(q)$, where now $q=1-\eta$ depends on the detector
efficiency.

\subsection{The $q >1$ case}

As we previously discussed, we can produce an NHH by considering $q=0$.
In this case, the dynamics of the system is completely determined by $\hat{H}_{\rm eff}$.
Thus, studying $\LL_H(0)$ we can infer if the effect of the NHH is to create or destroy an EP.

However, we cannot know in this formalism what is the effect of \emph{only} the quantum jumps on the system.
To do that, one should consider the $q\to \infty$ limit, where the NHH can never act on the system.
From the previous discussion it is clear that, to ensure a correct interpretation of the hybrid-Liouvillian in terms of post-selected trajectories, we need $q\in [0,1]$.
Therefore, $\LL_H(q>1)$ cannot be obtained by simply considering a post-selection procedure.

Even if we cannot provide a clear physical interpretation to $\LL_H(q>1)$, we can still study what happens to the spectrum of $\LL_H(q\gg 1)$ mathematically.
In this limit, the overall evolution of the system is given by the quantum jump operator, and the fate of the EPs of $\LL_H(q\to \infty)$  tells us if quantum jumps
either favor or are detrimental for the emergence of EPs.

\begin{figure*}[p]
    \centering
    \includegraphics[width=0.88\linewidth]{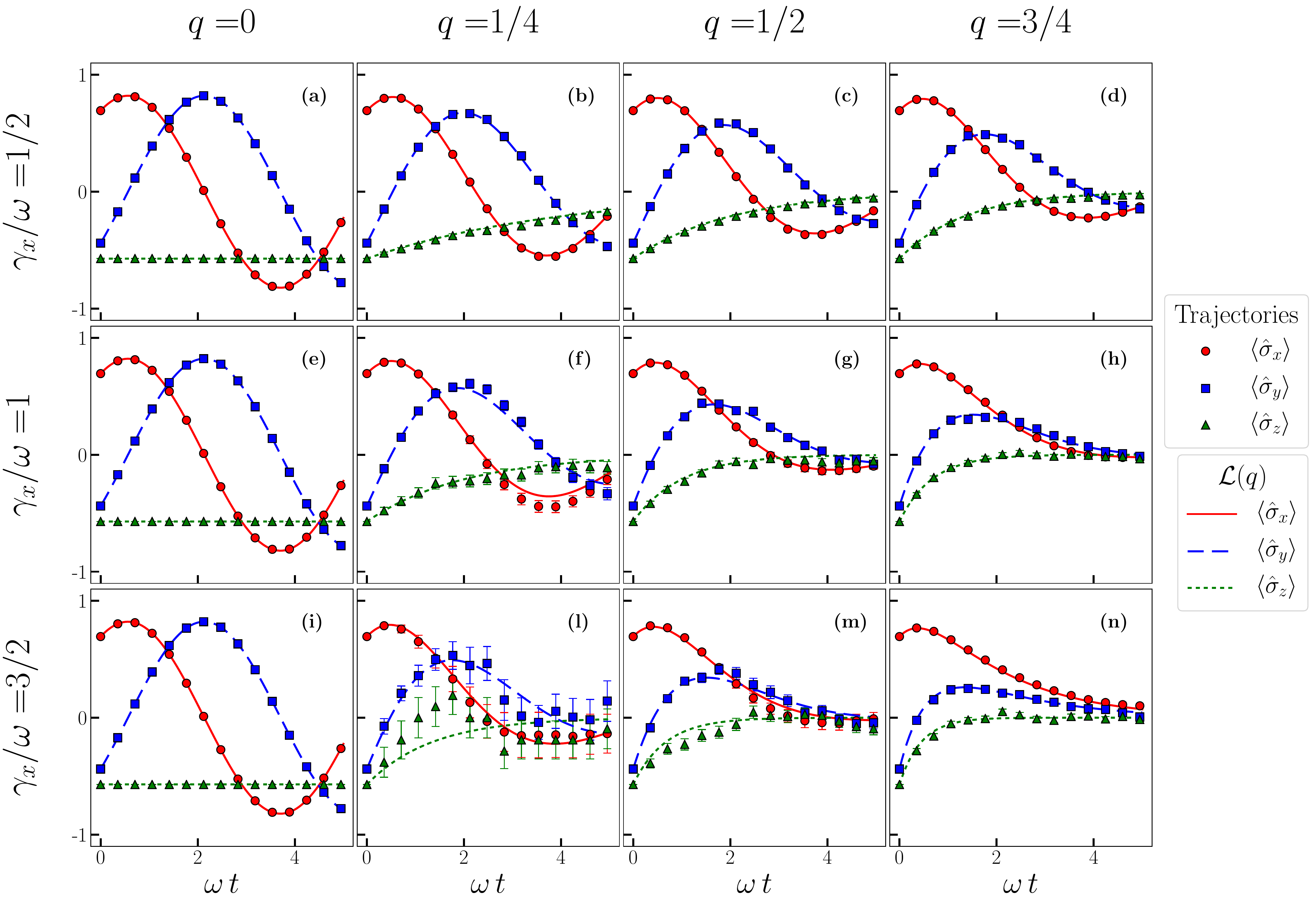}
\caption{Two-detector post-selected trajectories [c.f.
Fig.~\ref{fig:twodetectors} (a)] vs evolution using $\LL_H(q)$ for
the two-level system in Eq.~\eqref{Eq:hybrid_liouv_spin}: as a
function of time are plotted the expectation values of
$\hat{\sigma}_{x}$ (red curves and markers), $\hat{\sigma}_{y}$
(blue curves and markers), and $\hat{\sigma}_{z}$ (green curves
and markers).  The
markers represent the results obtained by the postselection of the trajectories, while the curves represent the results obtained via the hybrid
Liouvillian: [(a), (e),
 (i)]  $q=0$, [(b), (f), (l)]  $q=1/4$; [(c), (g),
(m)] $q=1/2$, and [(d), (h), (n)]  $q=3/4$; (a-d) $\gamma_{x}=\omega/2$, (e-h) 
$\gamma_{x}=\omega$, and (i-n) $\gamma_{x}=3 \omega/2$. The
initial state is $\ket{\psi}=\cos(\theta/2) \ket{\uparrow}+
\sin(\theta/2) e^{i \phi} \ket{\downarrow}$,  for
$\theta=\sqrt{3}\pi /2$ and $\phi=\sqrt{3} \pi$. 
The data have been obtained by averaging over
5000 trajectories per parameter set (see the details about the
algorithm in the main text).
The error bars $\Delta \braket{\hat{\sigma}_{i}} $ for a generic operator $\hat{\sigma}_{i}$ have been obtained by computing the standard error of the mean $\Delta \braket{\hat{\sigma}_{i}} = \sqrt{\braket{\hat{\sigma}_{i}  - \braket{\hat{\sigma}_{i}} }^2}/\sqrt{N-1}$, with $N$ representing the number of trajectories on which the average was taken.}
    \label{fig:post-trajvshybrid}
\end{figure*}
\begin{figure*}[p]
    \centering
    \includegraphics[width=0.88\linewidth]{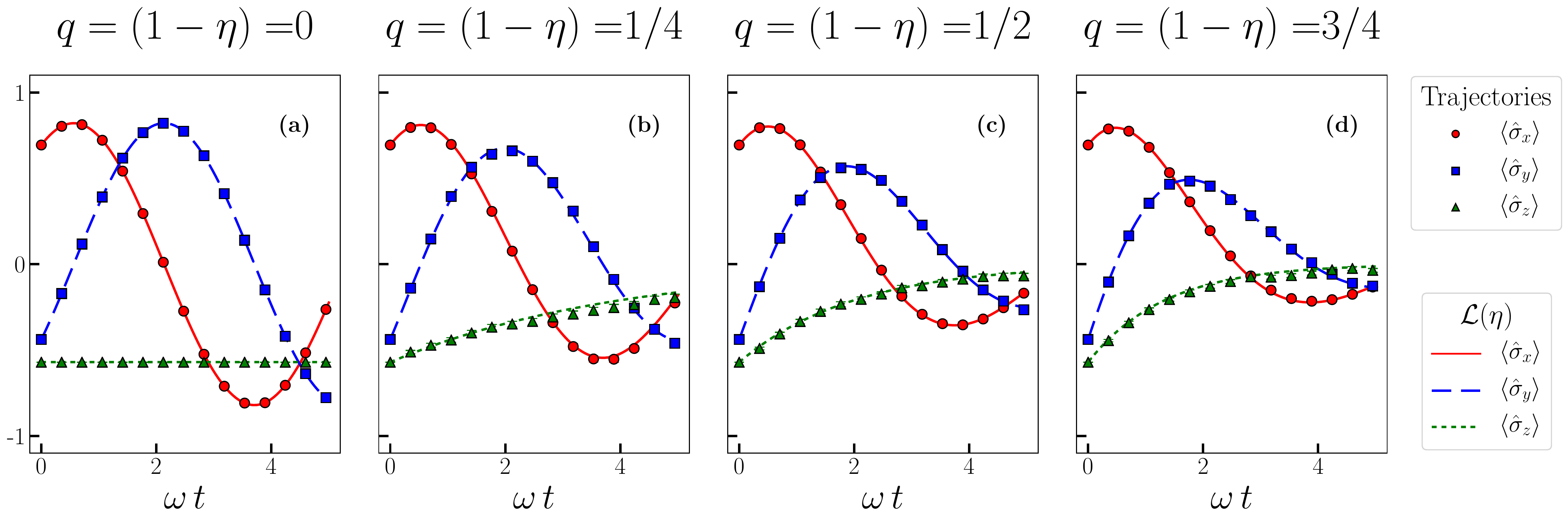}
    \caption{
Inefficient detector [c.f. Fig.~\ref{fig:twodetectors} (b)] vs
evolution using $\LL_H(q)=\LL_H(1-\eta)$ for the two-level system
in Eq.~\eqref{Eq:hybrid_liouv_spin} for $\gamma_{x}=\omega/2$. As a function of time, we
plot the expectation values of $\hat{\sigma}_{x}$ (red lines and
markers), $\hat{\sigma}_{y}$ (blue lines and markers), and
$\hat{\sigma}_{z}$ (green lines and markers).The markers represent the results obtained by the postselection of the trajectories, while the curves represent the results obtained via the hybrid
Liouvillian: (a) $q=0$
($\eta=1$), (b) $q=1/4$ ($\eta=3/4$), (c) $q=1/2$
($\eta=1/2$), and (d)  $q=3/4$ ($\eta=1/4$). The data have been obtained by averaging over
5000 trajectories per parameter set (see the details about the
algorithm in the main text) and the initial state is the same as in
Fig.~\ref{fig:post-trajvshybrid}. The error bars have been calculated as in Fig.~\ref{fig:post-trajvshybrid}.}
    \label{fig:inefficientdetector-vs-hybrid}
\end{figure*}

\section{Example 1: a model with a LEP without HEPs}
\label{Sec:Example_1}

In this Section, we consider a simplified version of the model
exhibiting LEPs but not HEPs given in Ref.~\cite{MingantiPRA19}.
We consider a spin-$1/2$, with Hamiltonian
\begin{equation}
\hat{H} = \frac{\omega}{2} \hat{\sigma}_z,
\end{equation}
which evolves under the action of the decay channel
$\hat\sigma_x$, i.e.,
\begin{equation}\label{Eq:Spin_liouvillian}
\LL \rhot= - i [\hat{H}, \rhot]
+\frac{\gamma_x}{2}\mathcal{D}[\hat\sigma_x],
\end{equation}
where $\hat\sigma_{x,\, z}$ are the Pauli matrices. Since this
master equation is invariant under the exchange $\hat{\sigma}_-
\to -\hat{\sigma}_-$, this model explicitly presents a
$\mathcal{Z}_2$ symmetry~\cite{AlbertPRA14,MingantPRA18_Spectral}.

As discussed in Ref.~\cite{MingantiPRA19}, the NHH structure is
trivial and cannot present any EPs, since
\begin{equation}
\hat{H}_{\rm eff}=\frac{\omega}{2} \sigma_{z} - i \gamma_{x} \mathbbm{1}
\end{equation}
is already diagonal, and its two eigenvalues are always different. The Liouvillian, instead, can present several interesting
properties. We have the eigenvalues
\begin{equation}\label{Eq:Spin_two_axes_vals}
\left\{\begin{split}
\lambda_0 &= 0,\\
\lambda_{1, \, 2} &= -\gamma _x \pm\Omega, \\
\lambda_3&=-2 \gamma_x,
\end{split}\right.
\end{equation}
and the corresponding eigenmatrices
\begin{equation}\label{Eq:Spin_two_axes_eigen}
\left\{\begin{split} \eig{0}&\propto \sss=\frac{1}{2 \gamma
    _x}\begin{pmatrix}
\gamma _x& 0 \\
0 & \gamma _x
\end{pmatrix},  \\
\eig{1, \, 2}&\propto \begin{pmatrix} 0 & -i
\omega \pm \Omega \\
\gamma _x & 0 \\
\end{pmatrix}, \\
\eig{3}&\propto\begin{pmatrix}
-1 & 0 \\
0 & 1 \\
\end{pmatrix},
\end{split}\right.
\end{equation}
where $\Omega= \sqrt{\gamma_x^2-\omega ^2}$ and $\sss$ is the
steady-state density matrix.
Therefore, this Liouvillian exhibits an EP for
$\gamma_{x}^{\rm EP}=\omega$.
For the model under consideration, we obtain a family of solutions for the generalized eigenmatrix $\hat{\tilde{\rho}}_1$ depending on one parameter $a$ [c.f. Eq.~\eqref{Eq:Jordan_chain}]:
\begin{equation}
\hat{\tilde{\rho}}_1 = \begin{pmatrix}
	0 & a\\
	 i\, a - i& 0
\end{pmatrix}.
\end{equation}

\subsection{Equivalence between postselection trajectories and hybrid Liouvillian}

Here, we show the equivalence between postselection and the
hybrid Liouvillian $\LL_H(q)$ stemming from
Eq.~\eqref{Eq:Spin_liouvillian}:
\begin{equation}\label{Eq:hybrid_liouv_spin}
\LL \rhot= - i [\hat{H}, \rhot] + q
\frac{\gamma_x}{2}\mathcal{D}[\hat\sigma_x] -
\frac{(1-q)\gamma_{x}}{2} \mathbbm{1} \bigcdot \, .
\end{equation}

\begin{figure*}[ht!]
    \centering
    \includegraphics[width=0.95\linewidth]{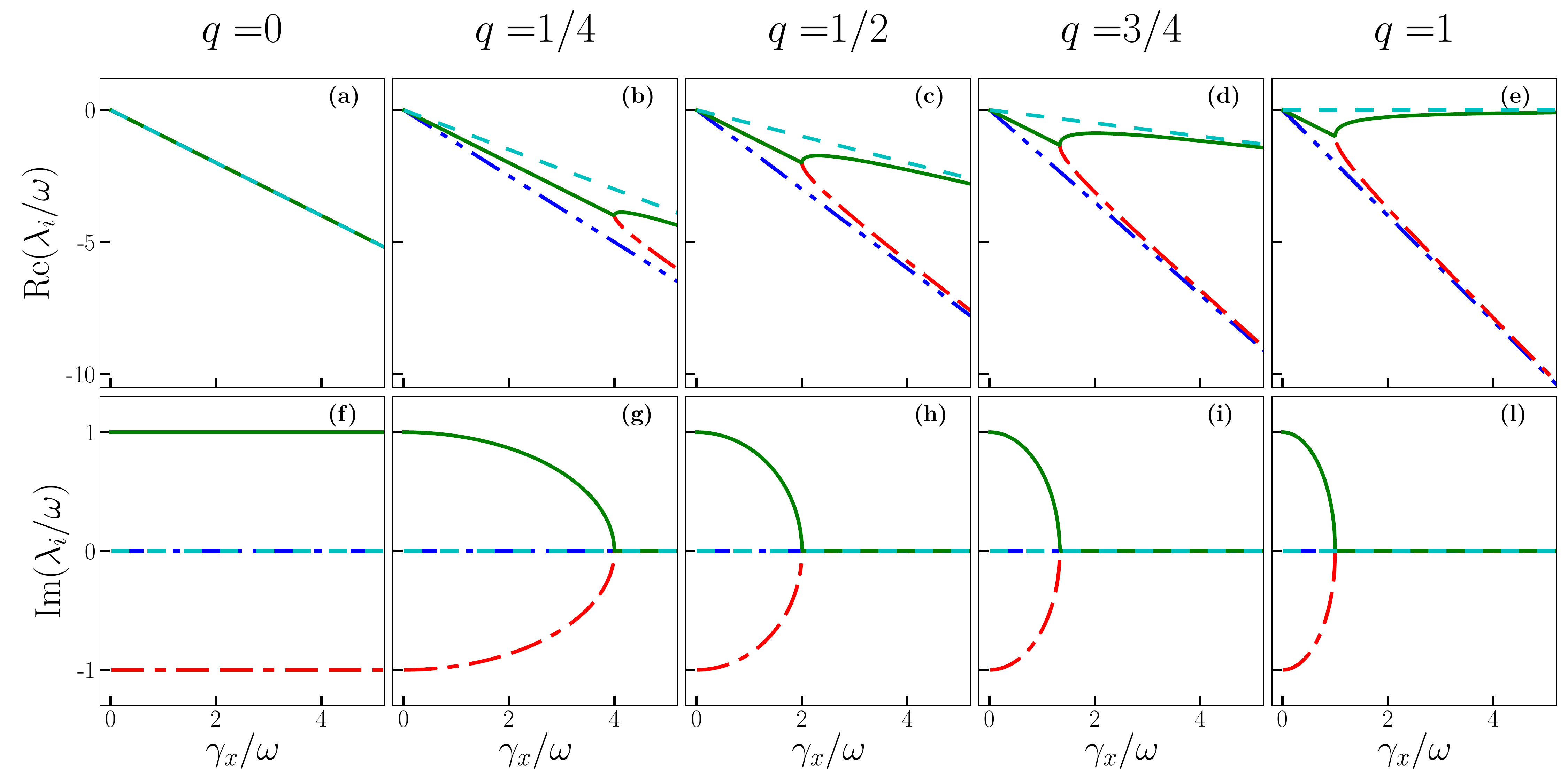}
    \caption{Spectrum of the hybrid-Liouvillian superoperator in Eq.~\eqref{Eq:hybrid_liouv_spin} as a function of the rescaled  dissipation rate $\gamma_{x}/\omega$.
         (a-e) The real part of the eigenvalues $\lambda_{i}$, and panels (f-l) the imaginary part.
[(a), (f)] $q=0$ (the NHH case), [(b), (g)]  $q=1/4$, [(c), (h)] $q=1/2$, [(d), (i)] $q=3/4$, and [(e), (l)] $q=1$ [i.e., diagonalization of the Liouvillian $\LL$ in Eq.~\eqref{Eq:Spin_liouvillian}].
    }
    \label{fig:spectrumpdf}
\end{figure*}

\subsubsection{Two-detector post-selected trajectories}

In Fig.~\ref{fig:post-trajvshybrid}, we study the time evolution
of a qubit both using $\LL_H(q)$ and post-selected trajectories
for the two-detector model [see also the sketch in
Fig.~\ref{fig:twodetectors} (a)]. The markers represent
the results obtained by averaging over 5000 single quantum
trajectories, from which we postselected only the evolutions
activating a chosen detector. The error bars are associated with
the statistical error due to the finite number of trajectories simulated. The curves, instead, have been obtained by evolving
the initial state via $\LL_H(q)$.

The algorithm to simulate this protocol for a number $N_{\rm
traj}$ of trajectories is the following: 
\begin{description}
    \item[Step 1] Simulate a quantum trajectory with two possible jump operators: $$\jump_1=\sqrt{q \gamma_{x}/2}\,\hat{\sigma}_{x}, \text{ \quad and \quad} \jump_2=\sqrt{(1-q)\gamma_{x}/2}\,\hat{\sigma}_{x};$$
    \item[Step 2] Once the simulation of one trajectory is completed, check the total number of jumps which took place in $\jump_2$, i.e., $N^{(2)}(t)$: if $N^{(2)}(t)=0$, save the trajectory, otherwise reject it;
    \item[Step 3] Once $N_{\rm traj}$ trajectories have been simulated, average only on the correct one.
\end{description}

Figures.~\ref{fig:post-trajvshybrid}(a), \ref{fig:post-trajvshybrid}(e), and \ref{fig:post-trajvshybrid}(i) (left column)  represent the results for $q=0$, that is, no
detection occurred, for $\gamma_{x}=\omega/2$,
$\gamma_{x}=\omega$, and $\gamma_{x}=3\omega/2$, respectively.
Correctly, the evolution is identical for the three different
values of $\gamma_{x}$, as it stems from the NHH. If we consider
now panels \ref{fig:post-trajvshybrid}(b)-\ref{fig:post-trajvshybrid}(d), which represent $\gamma_{x}=\omega/2$, we
observe a perfect agreement between the trajectories' behavior and
that of $\LL_H(q)$, demonstrating the validity of the previous
discussion. Note also that changing $q$ produces a sizable effect
on the system. If we increase the value of $\gamma_{x}$ [panels
(f-h), $\gamma_{x}=\omega$] there are some deviation from the
results of the hybrid Liouvillian. This noise is due to the
fact that, by increasing $\gamma_x$, the mean number of quantum
jumps increases and, therefore, fewer and fewer
trajectories which can be postselected.  This effect becomes
evident in  Figs.~\ref{fig:post-trajvshybrid}(i)-\ref{fig:post-trajvshybrid}(n), where $\gamma_x =3\omega/2$. In
particular, in Fig.~\ref{fig:post-trajvshybrid}(l), out of the 5000 trajectories which we
simulated, only 15 could be averaged. Moreover, the results in
Fig.~\ref{fig:post-trajvshybrid}(n) are far less noisy than those in Fig.~\ref{fig:post-trajvshybrid}(l): a higher $q$
means a smaller rejection rate, and therefore we can average over
a much higher number of trajectories.

\subsubsection{Imperfect detection}

Similarly to the previous case, in
Fig.~\eqref{fig:inefficientdetector-vs-hybrid} we consider now an
imperfect detector, similar to the one sketched in
Fig.~\ref{fig:twodetectors}(b). The detector efficiency $\eta$
represents the probability that when a quantum jump happens it is
detected. Thus, $\eta=1$ is a perfect detector, and all the
quantum jumps are detected. We simulate 5000 trajectories and we average only on those
which, according to our imperfect photodetector, had zero
quantum jumps.

The algorithm simulating this imperfect detection for a number
$N_{\rm traj}$ of trajectories is the following:
\begin{description}
    \item[Step 1] Simulate a quantum trajectory with one jump
    operator: $$\jump=\sqrt{\gamma_{x}/2} \,\hat{\sigma_{x}} ;$$
    \item[Step 2] Once the simulation of one trajectory is performed,
    we store those trajectories where all the quantum jumps which
    took place have not been detected, or no jump took place. To
    do that, we count the number $N_j$ of quantum jumps. We extract
    an array of $N_j$ random numbers $\{n_j\in [0,1]\}$, representing the
    aleatory nature of the imperfect detector in non-detecting the
    quantum jump. If, for all $j$, $n_j>\eta$ (or, equivalently, $n_j>1-q$) we save the trajectory. Otherwise reject it.
    \item[Step 3] Once $N_{\rm traj}$ trajectories have been simulated, average only on the correct ones.
\end{description}

Again, in Fig.~\eqref{fig:inefficientdetector-vs-hybrid} we see an
excellent agreement between the postselected averaged
quantum-trajectory (markers) and the evolution via operator
$\LL_H(q)=\LL_H(1-\eta)$ (curves). We have
therefore demonstrated the validity of our proposed protocol
and its physical meaning also in describing the physics of an
imperfect detector.

\subsection{Transition of NHH to $\LL$ and the appearance of a LEP}

Having demonstrated the validity of the physical interpretation of
$\LL_H(q)$,  we address now the question of the
emergence of a LEP in this model as a function of the control
parameter $q$.

As pointed out in Ref~\cite{MingantiPRA19}, the example studied in this
section does not present a HEP, but has a LEP. Therefore, we
can study the effect of the $q$ parameter on the emergence of
the EP.


For that purpose, let us first write explicitly the eigenvalues and eigenmatrices of this hybrid Liouvillian ${\cal
L}_{\rm H}(q)$ in Eq.~\eqref{Eq:hybrid_liouv_spin}. Its eigenvalues
and eigenmatrices read as follows
\begin{equation}\label{Eq:Spin_two_axes_vals_q}
\left\{\begin{split}
\lambda_{0,3} &= -\gamma_x(1\mp q),\\
\lambda_{1, \, 2} &= -\gamma _x \pm\Omega^{'}, \\
\end{split}\right.
\end{equation}
and
\begin{equation}\label{Eq:Spin_two_axes_eigen_h}
\eig{1, \, 2}\propto \begin{pmatrix} 0 & -i
\omega \pm \Omega^{'} \\
\gamma _x & 0 \\
\end{pmatrix},
\end{equation}
where
\begin{equation}\label{Omega1}
\Omega^{'}= \sqrt{q^2\gamma_x^2-\omega ^2},
\end{equation}
and the eigenmatrices $\eig{0,3}$ are the same as
in~\eqref{Eq:Spin_two_axes_vals}. In Figs.~\ref{fig:spectrumpdf}(a)~\ref{fig:spectrumpdf}(e)
we plot the real  and  in Figs.~\ref{fig:spectrumpdf}(f)~\ref{fig:spectrumpdf}(l) imaginary parts
of the spectrum of $\LL_H(q)$.

As Eq.~\eqref{Omega1} indicates, the EP of the hybrid Liouvillian
${\cal L}_{\rm H}(q)$ takes the form
\begin{equation}\label{gammax}
\gamma_x^{\rm EP}(q)=\omega/q.
\end{equation}
Thus, for $q=0$ there is no EP (equivalently, the EP is located at
infinity), as also shown in Figs.~\ref{fig:spectrumpdf}(a) and \ref{fig:spectrumpdf}(f). Indeed, also the diagonalization of $\hat{H}_{\rm eff}$ predicts no
coalescence of eigenvalues, as indicated by
Eq.~\eqref{Eq:Spin_two_axes_eigen_h}. On the contrary, for $q=1$
there is a LEP  [see also Figs.~\ref{fig:spectrumpdf}(e)~and~\ref{fig:spectrumpdf}(l)].
At the EP,  we can solve the Jordan chain relation for $\LL(q)$ in Eq.~\eqref{Eq:Jordan_chain}, obtaining the generalized eigenvector $\hat{\tilde{\rho}}_1$:
\begin{equation}
\hat{\tilde{\rho}}_1 = \begin{pmatrix}
0 & 1\\
0& 0
\end{pmatrix}.
\end{equation}

 By introducing a small $q$, we see the emergence of an EP,
but for a value of $\gamma_{x}$ which, in accordance with
Eq.~\eqref{gammax}, is much larger than $\gamma_{x}^{\rm EP}=\omega$ predicted by the Liouvillian theory [Fig.~\ref{fig:spectrumpdf}(b)]. As we increase
$q$, however, we observe that the position and the characteristics
of the EP become more and more similar to those of the LEP [c.f.
Figs.~\ref{fig:spectrumpdf}(c)~and~\ref{fig:spectrumpdf}(d) and Figs.~\ref{fig:spectrumpdf}(h)~and~\ref{fig:spectrumpdf}(i)].

These results confirm the interpretation of the LEPs provided in
Ref.~\cite{MingantiPRA19}. Indeed, it is the backaction of a
measurement apparatus on the system, induced by the quantum jumps,
that generates the EP~\cite{Barnett_BOOK_Info,
Haroche_BOOK_Quantum,Wiseman_BOOK_Quantum}. The projection of the
system on the eigenspace of its pointer states is attenuated by
the parameter $q$, thus a greater value of $\gamma_x$ is required
to observe the EP.

We conclude that, in this example, quantum jumps are the term responsible for the EP, while the $\hat{H}_{\rm eff}$ tends to destroy it.

\begin{figure*}
    \centering
    \includegraphics[width=0.95\linewidth]{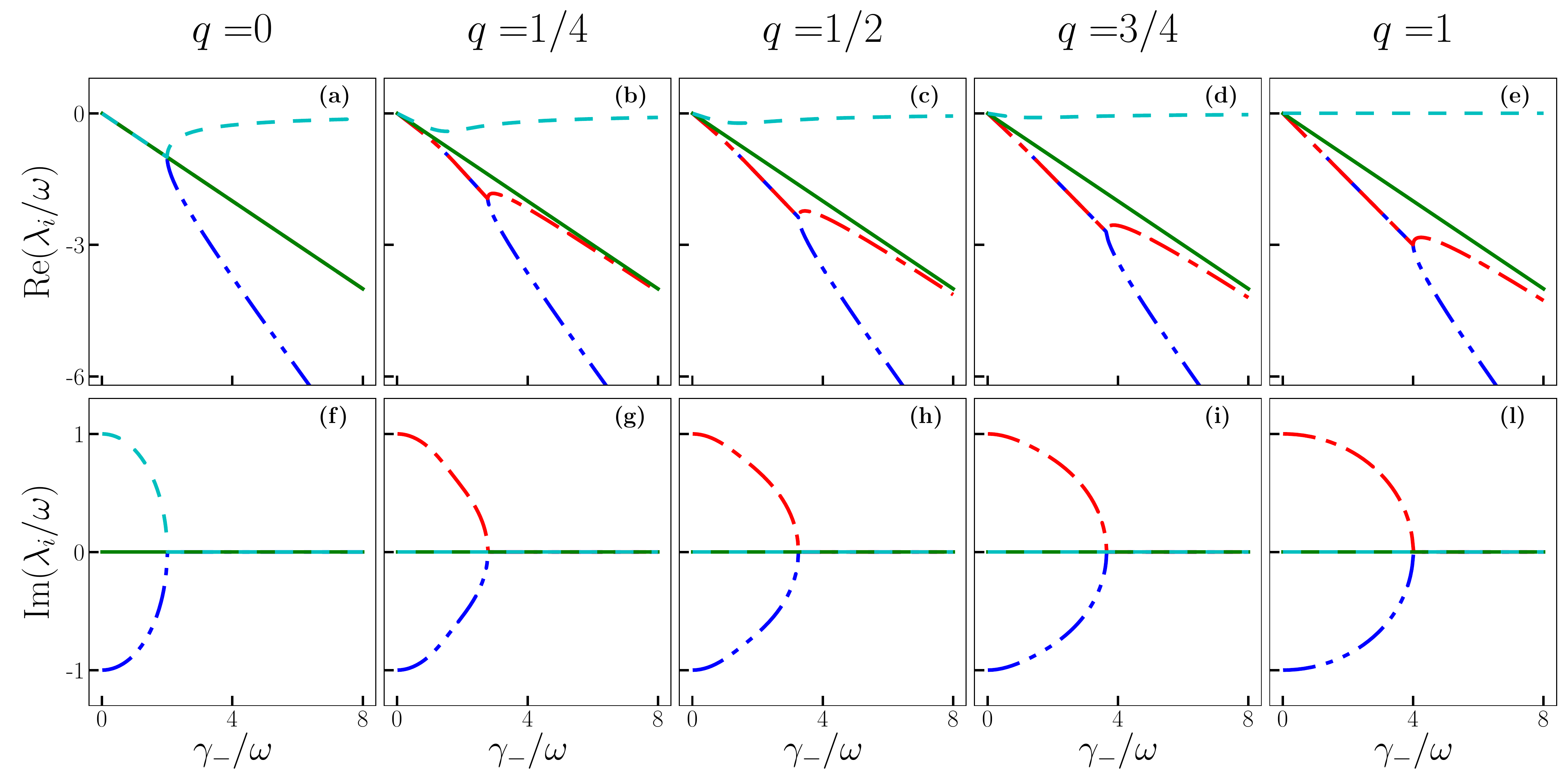}
    \caption{Spectrum of the hybrid-Liouvillian superoperator in Eq.~\eqref{Eq:Spin_hybrid_liouvillian_with_NNHEP} as a function of the rescaled  dissipation rate $\gamma_{-}/\omega$.
        (a-e) The real part of the eigenvalues and (f-l)  the imaginary one.
        [(a), (f)] $q=0$ (NHH case in Eq.~\eqref{Eq:NHH_with_EP}), [(b), (g)] $q=1/4$, [(c), (h)]  $q=1/2$, [(d), (i)] $q=3/4$, and [(e), (l)] $q=1$ [i.e., diagonalization of the Liouvillian in Eq.~\eqref{Eq:Spin_liouvillian_with_NNHEP}]. }
    \label{fig:hepandlep}
\end{figure*}

\begin{figure*}
    \centering
    \includegraphics[width=0.95\linewidth]{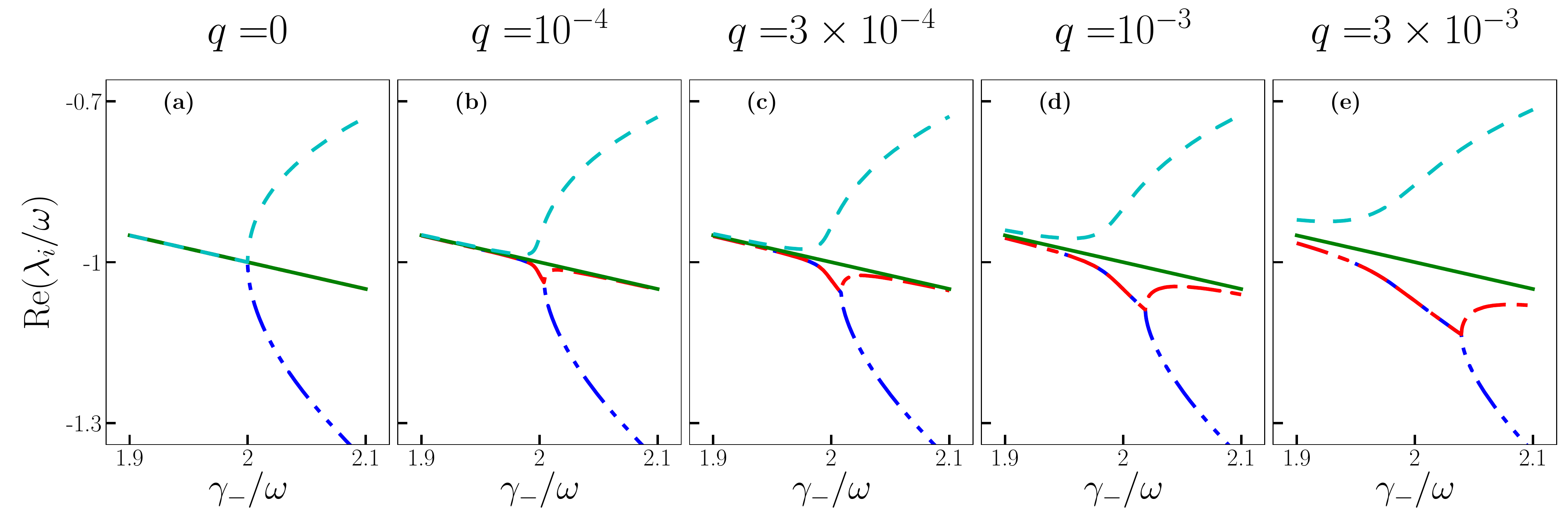}
    \caption{
Real part of the spectrum of the hybrid-Liouvillian
superoperator in Eq.~\eqref{Eq:Spin_hybrid_liouvillian_with_NNHEP}
as a function of the rescaled  dissipation rate
$\gamma_{-}/\omega$. The panels are for values of $q\gtrsim 0$,
showing how the inclusion of quantum jumps change the nature of
the EP.
    }
    \label{fig:hepandlep_zoom}
\end{figure*}

\section{ Example 2 of a model with inequivalent LEP and HEP}
\label{Sec:Example_2}

Let us now consider a model with HEPs and LEPs (as studied in
Ref.~\cite{MingantiPRA19}), where
\begin{equation}
\hat{H} = \frac{\omega}{2} \hat{\sigma}_x,
\end{equation}
which evolves under the action of the following Liouvillian
decaying channel:
\begin{equation}\label{Eq:Spin_liouvillian_with_NNHEP}
\LL \rhot= - i [\hat{H}, \rhot] + \frac{\gamma_- }{2}
\DD[\hat{\sigma}_-] \rhot.
\end{equation}
The NHH
\begin{equation}\label{Eq:NHH_with_EP}
\hat{H}_{\rm eff} = \frac{\omega}{2} \hat{\sigma}_x - i
\frac{\gamma_-}{2} \hat{\sigma}_+ \hat{\sigma}_-,
\end{equation}
which results from Eq.~\eqref{Eq:Spin_liouvillian_with_NNHEP} if
we ignore the quantum jump term in $\DD[\hat{\sigma}_-]$, has the
following eigenvalues:
\begin{equation}\label{Eq:spectrum_NHH_spin_with_HEP}
h_{1, \, 2} =\frac{1}{4} \left(-i \gamma_{-} \mp \zeta \right),
\end{equation}
and eigenvectors:
\begin{equation}\label{Eq:Ham_eigen_EP}
\ket{\phi_{1, \, 2}} \propto \left[i \gamma_- \mp \zeta \,, \quad
-2 \omega\right],
\end{equation}
where $\zeta=\sqrt{4 \omega^2-\gamma_{-}^2}$. Thus, this model has
a HEP for $\gamma_-=2\omega$,  admitting a family of generalized eigenvector depending on a single parameter $a$:
\begin{equation}
\ket{\tilde{\phi}_1}=\left[a, i  (4 + a) \right].
\end{equation}

The Liouvillian eigenvalues are instead
\begin{equation}\label{Eq:Liouvillian_spectrum_spin_system_with_HEP}
\begin{split}
\lambda_0&=0, \\
\lambda_1&=-\frac{\gamma _-}{2},\\
\lambda_{2, \, 3}&=-\frac{3}{4} \gamma_- \pm \beta/4 , \\
\end{split}
\end{equation}
and the eigenmatrices are
\begin{equation}\label{Eq:Eigenmatrices_Liouvillian_driven}
\begin{split}
\eig{0} & \propto \sss = \frac{1}{\gamma_{-}^2 + 2\omega^2}\begin{pmatrix}  \gamma _-^2+\omega ^2 & i \gamma _- \omega  \\
-i \gamma _- \omega  & \omega ^2
\end{pmatrix}, \\
\eig{1} & \propto
\begin{pmatrix}
0 & 1 \\
1 & 0 \\
\end{pmatrix}, \\
\eig{2, \, 3} & \propto \begin{pmatrix}
-\gamma _- \pm \beta & 4 i \omega  \\
- 4 i \omega  & \gamma _- \mp \beta
\end{pmatrix},
\end{split}
\end{equation}
where $\beta=\sqrt{\gamma _-^2-16\omega ^2}$. Hence, there is a
LEP for $\gamma_-=4\omega$.
The associated generalized eigenvector is:
\begin{equation}
\hat{\tilde{\rho}}_2 = 4  \, \begin{pmatrix} 1 & 0 \\
0 & -1
\end{pmatrix}.
\end{equation}

\subsection{Transition from the HEP to the LEP}

We study the effect of the jump parameter $q$ using the
hybrid-Liouvillian
\begin{equation}\label{Eq:Spin_hybrid_liouvillian_with_NNHEP}
\LL = - i [\hat{H}, \bigcdot] + q \frac{\gamma_- }{2}
\DD[\hat{\sigma}_-]\bigcdot - \frac{(1-q) \gamma_-}{2}
\frac{\hat{\sigma}_+ \hat{\sigma}_- \bigcdot +
\bigcdot\hat{\sigma}_+\hat{\sigma}_-}{2}.
\end{equation}

The analytical computation of the spectrum of the hybrid
Liouvillian in Eq.~\eqref{Eq:Spin_hybrid_liouvillian_with_NNHEP}
becomes more involved. Both its eigenvalue $\lambda_1$ and the
corresponding eigenmatrix $\eig{1}$ coincide with those given in
Eqs.~\eqref{Eq:Liouvillian_spectrum_spin_system_with_HEP} and
\eqref{Eq:Eigenmatrices_Liouvillian_driven}, respectively. On the
other hand, the remaining three eigenvalues $\lambda_{0,2,3}$
are the solutions of a third-order polynomial. Their
explicit form, along with their corresponding eigenmatrices, is
given in the Appendix~\ref{App:eigvals}.

Our analysis reveals an explicit dependence of the EP on
the parameter $q$, namely,
\begin{equation}\label{EP2h}
\gamma_-(q)=\sqrt{2}f^{-\frac{1}{2}}\left(3f^2+3q^2+2f\right)^{\frac{1}{2}}\omega,
\end{equation}
where
\begin{equation}
f = q^{\frac{2}{3}}\left(1+\sqrt{1-q^2}\right)^{\frac{1}{3}}.
\end{equation}
Clearly, as it follows from Eq.~\eqref{EP2h}, for $q=0$ ($q=1$)
one recovers the corresponding HEP (LEP).
In this case  we cannot analytically recover the form of the generalized eigenmatrix due to the presence of the $f$ terms.

We study the passage from the HEP to the LEP in
Fig.~\ref{fig:hepandlep}. We plot the real [Figs.\ref{fig:hepandlep}(a)-\ref{fig:hepandlep}(e)] and
imaginary [Figs.~\ref{fig:hepandlep}(f)-(l)] parts of the spectrum of $\LL_H(q)$ from
a standard NHH description [$q=0$ in Figs.\ref{fig:hepandlep}(a)~and~\ref{fig:hepandlep}(f)] to a
fully Liouvillian approach [$q=1$ in Figs.\ref{fig:hepandlep}(e)~and~\ref{fig:hepandlep}(l)]. By
increasing $q$ one moves away from the HEP picture, recovering the
spectral features resembling those of the LEP.

However, we notice that the mechanism which led to the EP is
very different in the Hamiltonian and Liouvillian cases. When
$q=0$, at the EP all the four different eigenvalues are coinciding.
Two of the corresponding eigenmatrices are associated with those of the NHH, while another
one is doubly degenerate [see the green solid curve in Fig.~\ref{fig:hepandlep}(a)]. Thus, while from a Hamiltonian perspective there is a second-order EP [c.f. Eq.~\eqref{Eq:Ham_eigen_EP}],
according to ${\cal L}_{H}(0)$ there is a third order EP. Indeed, as it stems from Figs.~\ref{fig:hepandlep}(a) and
\ref{fig:hepandlep}(f), the blue dotted-dashed curve and the light-blue dashed
are those producing the EP. 
The presence of an EP of higher order is not surprising since ${\cal L}_{H}(0)$ captures also the dynamics
of mixed states, which is impossible for the NHH.

In Figs. \ref{fig:hepandlep}(b) and \ref{fig:hepandlep}(g) we see that, even if the blue
dotted-dashed curve and the light-blue dashed curves change very
little, the EP is no more associated with the
coalescence of the states corresponding to those two. Indeed, the
blue dotted-dashed and the red curves show the generation of the
EP.
This abrupt change can be interpreted as the non-analyticity associated with the passage from a EP or order 3 to one of order 2 \cite{WiersigarXiv2020}.
In this regard, the effect of an imperfect post-selection can be extremely relevant in these systems.

To better grasp the effect of quantum jumps on this system, in
Fig.~\ref{fig:hepandlep_zoom} we plot the real part of the
Liouvillian spectrum for $q\gtrsim 0$. As we notice from Fig.~\ref{fig:hepandlep_zoom}(b), the introduction of a very small imperfection in the photodetector changes profoundly the
nature of the EP. In this limit, the states described by
$\LL_{H}(q)$ are almost identical to those described by
$\LL_{H}(0)$. Nevertheless, we may argue that in actual physical systems
the introduction of this minimal noise in the photodetector
counting is sufficient to distinguish between the three states
which previously coincided at the EP. Two will still produce
an EP for slightly shifted parameters. The other eigenmatrix,
instead, is pushed away from the spectral degeneracy.

We can interpret this result as a consequence of the exceptional sensitivity of the HEP to the presence of quantum noise.

Finally, for $q\to \infty$ the EP disappears [c.f. Eqs.~\eqref{A1},~\eqref{A2}, and~\eqref{A3}].
In this sense, we can say that in this model the presence of quantum jumps is detrimental to the emergence of an EP.

\section{Conclusions}

We studied the transition between two types of EPs:
semiclassical EPs (i.e., HEPs), which are degeneracies of
effective non-Hermitian Hamiltonians, and truly quantum
EPs (i.e., LEPs), which are degeneracies of a Liouvillian
superoperator corresponding to a Lindblad master equation,
as recently introduced in Ref.~\cite{MingantiPRA19} and applied in
Refs.~\cite{ArkhipovPRA20,HuangarXiv20}. We emphasize that the inclusion of quantum jumps in the
evolution of a quantum system makes, in general, LEPs fundamentally different from HEPs, as we have proved in Ref.~\cite{MingantiPRA19}.

n the present paper, we have addressed the question of
the relation of HEPs and LEPs based on the postselection
of quantum trajectories (quantum jumps) and their classicalto-quantum correspondence. This interpretation has partially
been inspired by a very recent experimental work~\cite{NaghilooNatPhys19} reporting
quantum state tomography of a single dissipative qubit in the
vicinity of its EP. This experiment was based on a \emph{post-selection} on a three-level superconducting circuit.

Here, we have applied the idea of postselection to propose
a hybrid-Liouvillian formalism based on a modified Liouvillian superoperator being a function of a quantum-jump
parameter $q$.
This approach directly shows the transition of
a LEP to a HEP via a proper postselection on quantum jumps
(or quantum trajectories) as schematically shown Fig.~\ref{fig:twodetectors}(a). Indeed, our formalism describes in particular (i) an NHH,
when one postselects only those trajectories without quantum
jumps (i.e., corresponding to the quantum jump parameter $q=0$),  and (ii) a true Liouvillian, including quantum jumps,
when one does not perform postselection (i.e., when $q=1$). Clearly, this formalism can describe also intermediate cases
for any $q\in[0,1]$, when we postselect a specific fraction of
trajectories.

Moreover, our approach allowed us to interpret postselection in an operational way based on finite-efficiency detectors.
Indeed, in addition to the analysis of the hybrid Liouvillian
in terms of the postselection of quantum trajectories, we
have also discussed its relation to inefficient photodetectors
corresponding to the case when a quantum jump occurs but
the detector does not signal it, as schematically shown in
Fig.~\ref{fig:twodetectors}(b).

We discussed two pedagogical examples showing the application of our general hybrid-Liouvillian approach. In our
first example, we analyzed a driven dissipative qubit model
exhibiting a LEP but without HEPs. In our second example,
we considered a qubit presenting a LEP and a HEP, which
occur for a different combination of parameters. The latter
example explicitly shows the transition of a LEP to a HEP
as a function of the quantum-jump parameter $q$.

Note that our examples show a double effect of quantum
jumps in creating and destroying EPs. Specifically, example 1
shows that a LEP [$\LL_{H}(q)$ for $q=1$] can be created solely by
quantum jumps, and no HEP  [$\LL_{H}(q)$ for $q=0$]  is generated in
this case. Contrary to this quantum-jump-induced EP, example
2 demonstrates a quantum-jump-destroyed EP, i.e., an NHH
can create a HEP, which would disappear for $q=\infty$ (where
quantum jumps would be the only process taking place). We
stress that for $q=1$  a LEP is observed, but its characteristics
are profoundly different from those of the HEP. This analysis
on the nature of HEPs and LEPs and on the roles played by the $\hat{H}_{\rm eff}$ and the quantum jumps could not have been performed
using the standard Liouvillian or NHH alone.

The advantages of EPs for applications remain a very
active topic of research
\cite{Wiersig2014,Zhang2015,Wiersig2016,Ren2017,KuoPRA20,Chen2017,Hodaei2017,ChenNat2018,LiuPRL16,Langbein2018,Lau2018,Mortensen2018,Wolff2019,Zhang2018,Chen2019,Langbein2018} and correctly modeling
noise and quantum jumps is fundamental to correctly address
the question of, e.g., EP sensitivity. We believe that our
work, showing explicitly the operational interpretation and
the relation between classical and quantum EPs in terms of
postselection and/or inefficient detectors, can stimulate more
interest in experimental demonstrations of LEPs and their
potential quantum applications, pointing out analogies and
differences with respect to those studied for semiclassical
HEPs.

Concerning the hybrid Liouvillian, it could be analyzed
in the context of decoherence-free subspaces, in particular
in relation to non-Hermitian dynamics in the presence of
dark states \cite{AlbertTHESIS}.
 In this case, the effect of the postselection
would be to change the eigenmatrices of the Liouvillian,
leaving the eigenvalues untouched. Moreover, it would be
interesting to try to generalize $\LL(q)$ to include, e.g., noisy
measurement instrument and time-dependent Hamiltonian or
jump operators, or to more general types of quantum maps,
such as jumptime unraveling of a quantum system\cite{Garrahan_JPA_2009,GneitingarXiv20}, or in connection to dynamical phase transitions \cite{MacieszczakPRA16}.
 Such
a hybrid Liouvillian may be also be studied in the context
of critical phenomena, where the $q$ parameter may change
the spectral properties normally associated with multistability
and metastability, multimodality, and critical slowing down \cite{MingantPRA18_Spectral,VicentiniPRA18,RotaNJP18,JinPRB18,LandaPRL20,MacieszczakPRL16,RosePRE16}.

\section*{Acknowledgments}
The authors acknowledge the discussion with M. Schir\'{o} and
comments of Jan Wiersig. 
F.M. was supported by the FY2018 JSPS Postdoctoral Fellowship for
Research in Japan. 
A.M. and R.C. were
supported by the Polish National Science Centre (NCN) under the
Maestro Grant No. DEC-2019/34/A/ST2/00081. 
I.A. thanks the Grant Agency of the Czech Republic (Project No. 18-08874S) and the project no. CZ.02.1.01/0.0/0.0/16\textunderscore019/0000754 of the Ministry of Education, Youth and Sports of the Czech Republic.
F.N. is supported in part by: NTT Research,
Army Research Office (ARO) (Grant No. W911NF-18-1-0358),
Japan Science and Technology Agency (JST)
(via the CREST Grant No. JPMJCR1676),
Japan Society for the Promotion of Science (JSPS)
(JSPS-RFBR Grant No. 17-52-50023, and JSPS-FWO Grant No. VS.059.18N),
and the Grant No. FQXi-IAF19-06 from the Foundational Questions Institute Fund (FQXi), a donor advised fund of the Silicon Valley Community Foundation.

\appendix
\section{Eigenvalues and eigenmatrices of the hybrid  Liouvillian in Eq.~\eqref{Eq:Spin_hybrid_liouvillian_with_NNHEP}}\label{App:eigvals}
\subsection{Eigenvalues $\lambda_i$}
By solving the  eigenproblem of the hybrid  Liouvillian
${\cal L}_{\rm H}(q)$ in
Eq.~\eqref{Eq:Spin_hybrid_liouvillian_with_NNHEP}, one arrives at the
following eigenvalues $\lambda_i$:
\begin{eqnarray}\label{A1}
\lambda_0&=&-\frac{\gamma_-}{2}+2F_0, \nonumber \\
\lambda_1&=&-\frac{\gamma_-}{2}, \nonumber \\
\lambda_{2,3}&=&-\frac{\gamma_-}{2}-F_0\pm i\sqrt{3}(F_0-2D),
\end{eqnarray}
where
\begin{equation}\label{A2}
F_0 = \frac{1}{12D}\left[D^2+3(\gamma_-^2-4 \omega^2)\right],
\end{equation}
and
\begin{equation}\label{A3}
D=\left[54q\gamma_-\omega^2+3\sqrt{3}\sqrt{108q^2\gamma_-^2\omega^4-(\gamma^2-4\omega^2)^3}\right]^{\frac{1}{3}}.
\end{equation}

Importantly, due to dependence of the eigenvalues $\lambda_i$ on the parameter $q$ in Eq.~\eqref{A1}, their sorting  [$|\Re{\lambda_i}|\leq |\Re{\lambda_{i+1}}|, $c.f. the text below Eq.~\eqref{Eq:spectrum}] is nontrivial. 
Thus, the indices in  $\lambda_i$ are reshuffled as $q$ changes. 
Consequently, also the corresponding eigenmatrices $\hat\rho_i$, present below, will undergo the same permutation of indices.

\subsection{Eigenmatrices $\hat\rho_i$}
The corresponding eigenmatrices $\hat\rho_i$ of the given hybrid
Liouvillian are listed below.
\onecolumngrid
 \emph{Eigenmatrix $\eig{0}$.---} The elements of the
eigenmatrix $\eig{0}$ read as follows
\begin{equation}
\begin{split}
\rho_{00}^{(0)}=&\frac{-1}{6}(D^3-54q\gamma_-\omega^2)(3\gamma_--D)+\frac{3\gamma^3(\gamma_--D)}{2} 
+\gamma^2\left[D^2-\omega^2(27q+12)\right]+3\gamma_-\omega^2D(3q+2) 
+24\omega^2-\omega^2D^2,  \\
\rho_{01}^{(0)}=&i3\omega D^2(4F_0-\gamma_-(2q+1)),  \\
\rho_{10}^{(0)}=&-\rho_{01}^{(0)},  \\
\rho_{11}^{(0)}=&6D^2\left[4\gamma_-qF_0+\omega^2D\right].
\end{split}
\end{equation}

 \emph{Eigenmatrix $\eig{1}$.---} The elements of this
eigenmatrix coincide with that $\eig{1}$ presented in
Eq~\eqref{Eq:Eigenmatrices_Liouvillian_driven}.

 \emph{Eigenmatrix $\eig{2}$.---} The  eigenmatrix $\eig{2}$
has the elements:
\begin{eqnarray}\label{eig2}
\rho_{00}^{(2)}&=&4D^2(\gamma_-^2-\omega^2)-\frac{u_+(D^3-9\gamma^3+36\gamma\omega^2)D}{3} \nonumber +{u_-\left[\gamma_-D^3-16\omega^2\gamma_-^2-3(\gamma_-^2-4\omega^2)^2\right]}, \nonumber \\
\rho_{01}^{(2)}&=&-i\omega D\left[D^2u_-+6\gamma_-D(2q+1)+3u_+(\gamma_-^2-4\omega^2)\right], \nonumber \\
\rho_{10}^{(2)}&=&-\rho_{01}^{(2)}, \nonumber \\
\rho_{11}^{(2)}&=&-2D\left[u_-q\gamma_-D^2-6\omega^2D+3u_+q\gamma_-(\gamma^2-4\omega^2)\right], \nonumber \\
\end{eqnarray}
where $u_{\pm}=1\pm\sqrt{3}$.

 \emph{Eigenmatrix $\eig{3}$.}---
The elements of the eigenmatrix $\eig{3}$ are the same as in
Eq.~\eqref{eig2}, except the change of the sign in the
off-diagonal elements, i.e.,  $\rho^{(3)}_{01}=-\rho^{(2)}_{01}$
and $\rho^{(3)}_{10}=-\rho^{(2)}_{10}$.

\twocolumngrid




%

\end{document}